# Robust evidence for the stabilization of the premartensite phase in Ni-Mn-In magnetic shape memory alloys by chemical pressure

Anupam K. Singh[1], Sanjay Singh[1]*, B. Dutta[2], K. K. Dubey[1], Boby Joseph[3], R. Rawat[4], Dhananjai Pandey[1]

[1]School of Materials Science and Technology, Indian Institute of Technology (Banaras Hindu University) Varanasi-221005, India.

[2]Department of Materials Science and Engineering, Faculty of Mechanical, Maritime and Materials Engineering, Delft University of Technology, Mekelweg 2, 2628 CD Delft, The Netherlands

[3]Elettra-Sincrotrone Trieste, Strada Statale14, Km 163.5 in Area Science Park, Basovizza 34149, Italy

[4]UGC-DAE Consortium for Scientific Research, University Campus, Khandwa Road, Indore 452001, India

**Abstract**

The thermodynamic stability of the premartensite (PM) phase has been a topic of extensive investigation in shape memory alloys as it affects the main martensite phase transition and the related physical properties. In general, the PM phase is stable over a rather narrow temperature-composition range. We present here evidence for chemical pressure induced suppression of the main martensite transition and stabilization of the PM phase over a very wide temperature range from 300 K to ~5 K in a magnetic shape memory alloy (MSMA) $Ni_{50}Mn_{34}In_{16}$ using magnetic susceptibility, synchrotron X-ray powder diffraction (SXRPD) studies and first-principles calculations. The ac-susceptibility studies show a highly skewed and smeared peak around 300 K without any further transition up to the lowest temperature of our measurement (5 K) for ~5% Al substitution. The temperature evolution of the SXRPD patterns confirms the appearance of the PM phase related satellite peaks at $T \leq 300$ K without any splitting of the main austenite (220) peak showing preserved cubic symmetry. This is in marked contrast to the temperature evolution of the SXRPD patterns of the martensite phase of the Al free as well as ~3% Al substituted compositions where the austenite (220) peak shows a clear splitting due to Bain distortion/symmetry breaking transition. Our theoretical calculations support the experimental findings and reveal that the substitution at the In site by a smaller size atom, like Al, can stabilize the PM phase with preserved cubic symmetry. Our results demonstrate that Al-substituted Ni-Mn-In MSMAs provide an ideal platform for investigating the physics of various phenomena related to the PM state.

## I. INTRODUCTION

Large shape change under the influence of external stress and its recovery on annealing above a characteristic temperature is the key property of a class of functional materials named shape memory alloys (SMAs)[1-4]. The origin of the large recoverable shape change (strain) in SMAs is intimately linked with a diffusionless martensite phase transition in which a higher symmetry cubic austenite phase transforms to a lower symmetry martensite phase with tetragonal/orthorhombic/monoclinic Bain distortion[5-7]. The martensite transition in SMAs is a reversible transition, unlike in the steels where it is irreversible[8]. An interesting precursor or premartensite (PM) transition has been reported to precede the martensite phase transition in some of the SMAs[3, 9, 10]. This phenomenon and its role on the structure-property correlations in various SMAs has since been extensively investigated[3, 10-15]. The precursor PM phase occurs at an intermediate temperature range between high temperature austenite and low temperature martensite phase with preserved cubic symmetry of the austenite phase[3, 10-15].

In recent years, the appearance of the PM phase has received considerable attention in another class of SMAs, known as magnetic shape memory alloys (MSMAs)[16-24]. The advantage of the MSMAs over the SMAs is that the magnetic field-induced shape change is accompanied with much larger strain (MFIS) which can be recovered within the martensite phase itself without any annealing above the transition temperature[25, 26]. Also, the response time associated with the occurrence of the magnetic field-induced strain and its recovery is much faster than that in the conventional SMAs[25] opening the possibility for the development of a new class of multifunctional sensors and actuators based on application/removal of magnetic field without any temperature variation[25]. Besides the large MFIS[26-28], the MSMAs have received tremendous interest in recent past due to the observation of several other exciting phenomena like large magnetocaloric effect[29-33], giant magnetoresistance[34-37], anomalous thermal properties[38, 39], exchange bias effect[40, 41], spin glass[42], giant Hall effect[43] and anomalous Nernst effect[44], all of which have great potential for technological applications. The study of the precursor PM phase in the MSMAs is of crucial importance in relation to several exotic phenomena like strain glasses[45] and skyrmions[24] in MSMAs.

Among the various MSMAs, the crystal structure, phase stability and the magnetisation behaviour of the PM phase has been extensively investigated in the Ni-Mn-Ga system[16, 18, 21, 22, 46, 47]. For example, the temperature dependence of the dc magnetisation in some of these alloys shows a small dip/peak at the PM transition[18, 19, 22] temperature at which it changes by ~2% as compared to that of the austenite phase[22, 23]. In contrast, the martensite phase transition is accompanied with a huge change (>40%) in magnetisation[22, 23] due to large Bain distortion and therefore much higher magnetocrystalline anisotropy of the martensite phase[48]. The PM phase of the near stoichiometric $Ni_2MnGa$ is characterized by the appearance of very weak intensity satellite peaks in the diffraction pattern even as the cubic austenite phase peaks remain almost unaffected due to the absence of any discernible Bain distortion[21, 22]. As a result, the PM phase has been regarded as a micro-modulated precursor state with preserved cubic symmetry[18]. This is in contrast to the martensite phase in which, besides the appearance of the new satellite peaks, cubic austenite peaks also split due to significant Bain distortion signaling a symmetry breaking transition[21, 22]. Recently, we presented evidence for the thermodynamic stability of both the PM and the martensite phases using a detailed temperature dependent SXRPD study on $Ni_2MnGa$[22].

The intermediate PM phase has been a subject matter of investigation in a few Ga-free Ni-Mn-X (X = Sn and In) MSMAs[23, 24, 49] also. For example, in Co-doped Ni-Mn-Sn alloys, obtained after pressure annealing[50, 51], the appearance of the precursor PM phase has been attributed to enhanced magnetoelastic coupling[18, 19, 23, 52]. Chemical pressure, generated by substitution with smaller size atom, like Al at the In site in the Ni-Mn-In alloy composition, has also been reported to stabilize the PM phase over a modest temperature range of about 40 K[23]. The stabilization of the PM phase has also been reported in the conventional shape memory alloys like NiTi after partial substitution of Ni with Fe[3] and in $Ni_xAl_{100-x}$ for $x < 60$[10, 53], by chemical pressure tuning.

The foregoing results highlight the importance of chemical pressure-tuning of the transition temperature leading to the stabilization of the precursor PM phase and destabilization of the martensite phase. However, all these studies are mainly based on bulk magnetic measurements without any temperature dependent structural studies. Since the laboratory source X-ray power diffraction data often fails to capture the signatures of the PM phase[54], any structural confirmation of the chemical pressure-tuned PM phase in MSMAs requires high-resolution SXRPD data, which not only reveal the presence of characteristic weak satellite peaks due to its high signal to noise ratio but also its ability to capture the Bain distortion, if present, and hence the signature of the symmetry breaking transition, unambiguously due to the high peak to peak resolution[22, 46, 55]. In the present work, we have carried out a comprehensive study on the effect of the chemical pressure generated by Al substitution in place of In in $Ni_{50}Mn_{34}In_{16}$ MSMA using bulk magnetic measurements, temperature dependent high-resolution SXRPD studies and first-principles calculations. Our magnetic susceptibility studies on these alloys suggest that Al substitution in place of In destabilizes the martensite phase and stabilizes the PM phase over a wider temperature range. Our temperature dependent high resolution SXRPD studies on these alloys reveal Bain distortion in the martensite phase and its absence in the PM phase. The bulk magnetic and structural studies show that Al free $Ni_{50}Mn_{35}In_{15}$ MSMA exhibits only the martensite phase of 3*M* type in the monoclinic space group *P*2*/m* with significant Bain distortion[56]. However, in the ~3% Al substituted $Ni_{50}Mn_{34}In_{15.5}Al_{0.5}$ alloy, a PM phase, stable over a narrow temperature window of ~10 K below the ferromagnetic $T_C$ ~ 317 K, precedes the appearance of the martensite phase at lower temperatures. The PM to martensite phase transition is shown to be an isostructural phase transition as both the phases are of 3*M* type in the monoclinic *P*2*/m* space group and differ only in terms of the absence or otherwise of the Bain distortion. More significantly, we show that on increasing the Al content to ~5% (i.e., $Ni_{50}Mn_{34}In_{15.2}Al_{0.8}$), only the PM phase, without any Bain distortion, occurs over the entire temperature range (300 K to 5 K) below the PM transition temperature $T_{PM}$ ~ 300 K without any signature of the martensite phase transition either in the magnetic or the structural studies. We also show that the $T_{PM}$ decreases with increasing magnetic field and the satellite peaks of the PM phase disappear in the presence of external magnetic field confirming the magnetoelastic coupling in this alloy composition. Using first-principles calculations, we further investigate the energetics of alloys which provide atomistic insight into the stabilization of PM phase in Al substituted Ni-Mn-In MSMA. Our results demonstrate that Al substituted Ni-Mn-In MSMAs provide an ideal platform for investigating the physics of PM phase related phenomena in MSMAs.

**II. RESULTS AND DISCUSSION**

The details of sample preparation, measurements and analysis are given in the Supplemental Material[57]; also see Refs.[58-68]. Al-free $Ni_{50}Mn_{35}In_{15}$ (or $Ni_2Mn_{1.4}In_{0.6}$) MSMA[56] exhibits paramagnetic to ferromagnetic (FM) transition with $T_C \sim 315$ K, a first order austenite to martensite transition in the FM phase at $T_M \sim 295$ K with a characteristic thermal hysteresis in the temperature dependent magnetisation $M(T)$ plots for the field cooled (FC) and field cooled warming (FCW) protocols and another transition at $T_C^M \sim 150$ K, commonly attributed to competing FM and antiferromagnetic (AFM) interactions[69, 70], with bifurcation of the zero field cooled warming (ZFCW) and FC $M(T)$ plots in the dc magnetization studies[56]. While the FM $T_C$ of the base alloy ($Ni_{50}Mn_{35}In_{15}$) is known to be nearly unaffected by Al substitution[23, 36], the nature of transitions below $T_C$ changes rather drastically as a function of Al content. This can be seen from a comparison of magnetization data of the base alloy in Ref.[56] with those given in Fig. 1(a) and 1(b). For the ~3% Al substituted composition ($Ni_{50}Mn_{34}In_{15.5}Al_{0.5}$) one observes, two peaks at $T_{PM} \sim 311$ K and $T_M \sim 300$ K in real part of ac-susceptibility ($\chi'(T)$) plot shown in the inset of Fig. 1(a) corresponding to the premartensite and martensite transitions, respectively. Both the transitions exhibit characteristic thermal feature in $\chi'(T)$ plots for the FC and FCW protocols, shown in the main figure (Fig. 1(a)), suggesting their first order character. The nature of the two transitions shown in Fig. 1(a) are in broad agreement with those reported in a previous study[23, 49], even though the transition temperatures and behavior of $\chi'(T)$ are somewhat different, possibly due to a small fluctuations in the alloy composition[70-74].

We now proceed to correlate the two anomalies in $\chi'(T)$ of $Ni_{50}Mn_{34}In_{15.5}Al_{0.5}$ with premartensite and martensite phase transitions using structural studies. Fig. 2(a), (b) and (c) depict the SXRPD patterns of $Ni_{50}Mn_{34}In_{15.5}Al_{0.5}$ recorded at 400 K (> $T_C$), 310 K and 110 K, respectively. The emergence of new peaks in these SXRPD patterns at 310 and 110 K reveal structural changes related with the premartensite and martensite transitions, respectively. All the peaks in Fig. 2(a) can be indexed with the austenite cubic structure in the $Fm\bar{3}m$ space group, as confirmed by LeBail refinement, the details of which are given in the Fig. S1(a) (see Supplemental Material[57]). The cubic lattice parameter obtained after the refinement is found to be $a = 6.01009(6)$ Å. Further, the presence of the (111) and (200) Bragg reflections (see the inset (i) of Fig. S1(a); see Supplemental Material[57]) confirms the L2$_1$ ordering above the FM $T_C \sim 317$ K. At $T \sim 310$ K ($< T_{PM}$), new satellite peaks with very low intensities appear, as can be seen from the inset (i) of Fig. 2(b), which gives the SXRPD plot on a magnified scale for a limited 2θ range. All the peaks, including the satellite peaks, in this pattern are well accounted for by a 3*M* modulated monoclinic structure in the *P*2/*m* space group with $a = 4.3869(7)$ Å, $b = 5.6866(1)$ Å, $c = 13.0028(2)$ Å and $\beta = 93.695(3)°$ as can be seen from Fig. S1(b) (see Supplemental Material[57]) which gives the results of LeBail refinement for the SXRPD pattern at ~310 K. At this temperature, the "cubic" peaks do not show any splitting (see inset (ii) of Fig. 2(b)) confirming the appearance of the PM phase below $T_{PM}$ with preserved cubic symmetry without any discernible Bain distortion, similar to the PM phase of Ni$_2$MnGa[46,75]. On lowering the temperature further below $T_M$, the intensity of the existing satellite peaks increases considerably while the "cubic" peaks split into multiple peaks (see the inset of Fig. 2(c) due to significant Bain distortion, as expected for the martensite phase[56]. LeBail refinement for the 110 K SXRPD pattern confirms that all the peaks in the martensite phase are also accounted for using the monoclinic *P*2/*m* space group (Fig. S1(c); see Supplemental Material[57]). We also verified the stability of the martensite phase from 300 K down to 13 K using x-ray powder diffraction data obtained from an 18 kW Cu rotating anode based high resolution diffractometer fitted with a curved graphite crystal monochromator in the diffraction beam and a close-cycle He refrigerator-based low temperature attachment. The corresponding x-ray diffraction (XRD) patterns in the 20 to 100° 2θ-range are depicted in Fig. S2 (see Supplemental Material[57]). Thus, the SXRPD studies on the $Ni_{50}Mn_{34}In_{15.5}Al_{0.5}$ alloy above 300 K reveals cubic austenite to monoclinic (space group *P2/m*) PM phase transition at $T_{PM}$ without any Bain distortion, as revealed by the absence of any splitting of the "cubic" peaks while the SXRPD and rotating anode based laboratory source data reveal an isostructural PM to martensite phase transition with significant Bain distortion, as revealed by the splitting of the "cubic" peaks in the martensite phase.

On increasing the Al content from ~3% to ~5% (i.e., $Ni_{50}Mn_{34}In_{15.2}Al_{0.8}$ MSMA), the phase transition behaviour changes drastically. Fig. 1(b) shows temperature dependence of the real part of $\chi'(T)$ for ~5% Al substituted alloy, $Ni_{50}Mn_{34}In_{15.2}Al_{0.8}$. For the FC protocol, the paramagnetic to FM transition occurs at $T_C$ ~ 317 K with a sharp increase in the $\chi'(T)$, comparable to the $T_C$ of the base alloy as can be seen from a comparison of Fig. 1(b) with the temperature dependence of dc magnetization of the base alloy in Ref.[56]. On decreasing the temperature further, the rate of increase of $\chi'(T)$ decreases before peaking at the PM transition temperature $T_{PM}$ ~ 300K, as can be seen from the inset of Fig. 1(b). This peak is highly skewed and smeared out on the lower temperature side down to ~5 K. The gradually decreasing trend of $\chi'(T)$ below $T_{PM}$ is in marked contrast to its sharp drop in dc magnetisation of the base alloy[56] which is usually attributed to a very large magnetocrystalline anisotropy of the martensite phase[48]. More significantly, there is no signature of the second anomaly, seen in Fig. 1(a) for $Ni_{50}Mn_{34}In_{15.5}Al_{0.5}$, corresponding to the martensite transition in $Ni_{50}Mn_{34}In_{15.2}Al_{0.8}$.

The absence of martensite transition and stabilization of the PM phase indicated by the $\chi'(T)$ plot in Fig. 1(b) was confirmed by SXRPD studies at selected temperatures in the 400 to 100 K range and laboratory source XRD patterns at several temperatures in the 300 to 13 K range. Fig. 3(a) compares the SXRPD patterns of $Ni_{50}Mn_{34}In_{15.2}Al_{0.8}$ alloy at 400, 220 and 100 K. The insets in panels (i), (ii) and (iii) depict a magnified view of the profiles in the 2θ range 5.38º to 5.84º around the most intense Bragg peak. The inset of (ii) reveals the presence of satellite peaks at 220 K whose intensity increases on lowering the temperature to 100 K (see inset of (iii)). These satellite peaks are absent at 400 K, as can be seen in the inset of (i). The evolution of the SXRPD profiles as a function of temperature is shown in Figs. 3(b) and (c) at close temperature interval using a magnified (vertically zoomed) view of the intensity profile around the most intense cubic (220) peak. The most intense satellite appears around 300 K as shown with an arrow in Fig. 3(d), in agreement with the transition temperature $T_{PM}$ corresponding to the austenite to PM transition in the $\chi'(T)$ plot shown in Fig. 1(b). The intensity of the three prominent satellite peaks around the (220) cubic peak keeps growing below 300 K, as can be seen from Figs. 3(b) and (c). Further, the FWHM of the satellite peaks decreases (see Fig. 3(c)) with decreasing temperature suggesting that the correlation length or the domain size of the PM phase keeps growing below 300 K after its nucleation around $T_{PM}$=300 K. Moreover, the cubic (220) peak does not show any splitting down to the lowest temperature 100 K upto which we could collect the SXRPD patterns (see Fig. 3(e)). We verified the absence of splitting in this peak even below 100 K using the laboratory source X-ray powder diffraction data (see inset of Fig. 3(e)). The absence of any splitting of the (220) cubic peak, shown in the inset of Fig. 3(e), down to 13 K is in marked contrast to that shown in Fig. S2(b) (see Supplemental Material[57]) $Ni_{50}Mn_{34}In_{15.5}Al_{0.5}$ alloy. This confirms the stability of the PM phase down to 13 K with preserved cubic symmetry without any discernible Bain distortion. The absence of any splitting of the (220) austenite peak, the appearance of the satellite peaks at $T \leq T_{PM} \sim 300$ K and the absence of any peak in the $\chi'(T)$ plot in Fig. 1(b) below the $T_{PM}$ corresponding to the martensite transition clearly confirms the suppression of the martensite phase in $Ni_{50}Mn_{34}In_{15.2}Al_{0.8}$ MSMA and the stabilization of the PM phase in the entire temperature range below $T_{PM}$. These qualitative observations were verified by LeBail refinement as discussed below.

The LeBail refinement using the SXRPD pattern at 400 K (Fig.4(a)) for the $Ni_{50}Mn_{34}In_{15.2}Al_{0.8}$ alloy for the cubic austenite phase in the space group $Fm\bar{3}m$ confirmed that all the peaks in the SXRPD pattern could be indexed very well. The results of this refinement are shown in Fig. 4(a) which reveals an excellent fit between the observed and calculated profiles for the refined unit cell parameter $a = 6.0169(1)$ Å. The presence of (111) and (200) Bragg peaks shown in the inset of Fig. 4(a) confirms the ordered $L2_1$ cubic structure for the austenite phase[56]. Having confirmed the single-phase nature and $L2_1$ ordering in the cubic austenite phase of $Ni_{50}Mn_{34}In_{15.2}Al_{0.8}$, LeBail refinement using the SXRPD pattern was carried out at 100 K (i.e., well below $T_{PM}$) for the 3$M$ modulated monoclinic structure in the space group $P2/m$, similar to that for the $Ni_{50}Mn_{34}In_{15.5}Al_{0.5}$ composition. The observed, calculated and difference profiles so obtained, shown in Fig. 4(b), reveal excellent fit for the 3$M$ modulated monoclinic structure. The inset of Fig. 4(b) depicts an enlarged view of the LeBail fit around the most intense Bragg peak. The satellite peaks corresponding to the PM phase are marked as 'PM' along with their indices in this inset. The refined lattice parameters obtained after refinements are ($a = 4.3823(2)$ Å, $b = 5.6480(4)$ Å, $c = 12.9754(4)$ Å, $\beta = 93.755(3)°$) for the PM structure of $Ni_{50}Mn_{34}In_{15.2}Al_{0.8}$ at 100 K. These parameters correspond to the 3$M$ modulated monoclinic structure as per the convention used in the literature for the modulated structures in MSMAs[21, 76-80].

The magnetoelastic coupling has been suggested as one of the factors for the stabilization of the PM phase in MSMAs[18, 23, 50, 81]. The effect of magnetoelastic coupling is manifested through the variation of $T_{PM}$ with magnetic field[18, 19, 23, 50]. Therefore, to investigate the possibility of magnetoelastic coupling in $Ni_{50}Mn_{34}In_{15.2}Al_{0.8}$ MSMA, we investigated the temperature dependence of the dc magnetisation as a function of temperature ($M(T)$) at different magnetic fields (100 Oe, 200 Oe, 500 Oe, 800 Oe, and 1500 Oe) during warming cycle on zero field cooled sample (ZFCW protocol) shown in Fig. 4(c). The enlarged view of Fig. 4(c) around the diffuse peak in the $M(T)$ is shown in Fig. 4(d) where the peak temperatures are marked by arrows. The variation of the peak temperatures in $M(T)$ with field, shown in Fig. 4(e), reveals that $T_{PM}$ shifts towards the lower temperature side linearly with increasing magnetic field. This has been attributed to magnetoelastic coupling[18, 19, 23, 50]. We verified the magnetoelastic coupling in $Ni_{50}Mn_{34}In_{15.2}Al_{0.8}$ further by recording high-resolution SXRPD patterns without and in the presence of magnetic field (0 Oe and 2500 Oe) at 294 K. The results of such measurements are shown in Fig. 4(f) around the most intense Bragg peak while the SXRPD pattern in full 2θ-range is shown in Fig. S4 (Fig. S4; see Supplemental Material[57]). It is evident from Fig. 4(f) that the satellite peaks related to the PM phase (marked as 'PM' for the zero-field pattern) disappear completely at 2500 Oe. This confirms that the lattice and spin degrees of freedom are not only coupled but also play a crucial role on the stability of the PM phase of $Ni_{50}Mn_{34}In_{15.2}Al_{0.8}$ MSMA. It is worth mentioning here that the satellite peaks related to the PM phase are better resolved in Fig. 4(f) than Fig. 3(b) due to better peak to peak resolution for the lower energy (25 keV) beam used for the former in contrast to 60 keV beam used for the latter.

In order to obtain microscopic insight into the role of Al substitution on the stability of the PM phase of Ni-Mn-In MSMA, we calculated total energies of both PM and austenite phases using first-principles simulations for three off-stoichiometric compositions, i.e., $Ni_{50}Mn_{33.3}In_{16.7}$, $Ni_{50}Mn_{33.3}In_{12.5}Al_{4.2}$ and $Ni_{50}Mn_{33.3}Al_{16.7}$ and the results so obtained are presented in Fig.5. Excess Mn atoms (i.e., above 25% Mn content) in our simulation cells (shown in insets (ii), (iii) of Fig. 5(a) and Fig. 5(b)) occupy the In/Al sub-lattice and couple ferromagnetically to the host Mn atoms in the two In containing alloys. The interaction between host and excess Mn atoms is, however, of antiferromagnetic nature in $Ni_{50}Mn_{33.3}Al_{16.7}$. We have considered different chemical arrangements of excess Mn, In and Al on the In/Al sub-lattice in the present study and selected the atomic configuration with the lowest total energy for both the PM and the austenite phases. Our calculations indicate that the modulated PM phase is not stable for the Al-free $Ni_{50}Mn_{33.3}In_{16.7}$ MSMA (results not shown here). During the process of ionic relaxation, initially chosen atomic modulation for the modulated PM phase, as discussed in Ref.[82], disappears and the structure becomes identical to that of the non-modulated austenite phase. We repeated the procedure several times with different values of amplitude of atomic modulation but always obtained the non-modulated austenite phase after performing ionic relaxation. The Al substitution in place of In tends to stabilize the modulated PM phase as shown in the inset (i) of Fig. 5(a). For the $Ni_{50}Mn_{33.3}In_{12.5}Al_{4.2}$ MSMA, the total energy at 0 K for the PM phase around its equilibrium volume is slightly lower (approximately 0.1 meV/atom) than that of the austenite phase. This suggests that Al substitution at the In site facilitates the stabilization of the modulated PM phase in $Ni_{50}Mn_{33.3}In_{12.5}Al_{4.2}$, albeit with tiny amplitudes of atomic modulations (inset (iii) of Fig. 5(a)) which nevertheless accounts for the small energy difference between the austenite and the PM phases for this alloy composition. In order to unambiguously establish the role of Al in stabilizing the modulated PM phase, we calculated the total energies of both the phases for an alloy with only Al in place of In, i.e., $Ni_{50}Mn_{33.3}Al_{16.7}$. The inset (i) of Fig. 5(b) shows that the total energy at 0 K for the PM phase around its equilibrium volume is lower than that of the austenite phase by more than 2 meV/atom for this In free alloy. The amplitudes of the atomic modulations in the PM phase are also substantially larger in this case, as shown in the inset (iii) of Fig. 5(b). All these suggest that the stability of the modulated PM phase increases with increasing Al content in place of In in Ni-Mn-In based MSMAs. It is important to mention that the temperature-dependent excitations, e.g., lattice vibrations, magnetic excitations etc., are likely to alter the relative stability of the austenite phase with respect to that of the PM phase at higher temperatures[83]. Nevertheless, at low temperatures, our calculations predict PM phase to be more stable than the austenite phase for Al-substituted Ni-Mn-In based MSMAs.

Moreover, we note that the stability of the PM phase in $Ni_{50}Mn_{33.3}Al_{16.7}$ is strongly dependent on the above-mentioned antiferromagnetic interaction between host and the excess Mn atoms. For an artificially chosen ferromagnetic state (Fig. S7; see Supplemental Material[57]) with all Mn spins aligned in the same direction, atomic modulations in the PM phase almost disappear resulting in negligible energy difference between the PM and the austenite phase. It may therefore be speculated that the origin of the missing PM phase in $Ni_{50}Mn_{33.3}In_{16.7}$ lies in the ferromagnetic interaction between the host and the excess Mn atoms in this alloy. The substitution of Al in place of In in $Ni_{50}Mn_{33.3}In_{16.7}$, however, tends to change the nature of this magnetic interaction from ferromagnetic to antiferromagnetic, which in turn facilitates the stabilization of the PM phase. Our calculations, thus, indicate a strong influence of magnetism on the stability of the PM phase which corroborates the experimental finding for the absence of the PM phase in the presence of a modest external magnetic field (Fig. 4 (f)).

## III. CONCLUSIONS

To conclude, we presented here evidence for chemical pressure induced suppression of the main martensite transition and stabilization of the PM phase over a very wide temperature range from 300 K to ~5 K in a magnetic shape memory alloy $Ni_{50}Mn_{34}In_{16}$ using magnetic susceptibility, synchrotron x-ray powder diffraction studies and first-principles calculations. The ac-susceptibility studies show that the stability of the martensite phase is suppressed while that of the premartesnite phase enhanced with increasing Al content in place of In in $Ni_{50}Mn_{34}In_{16}$. The temperature evolution of the SXRPD patterns provides robust evidence for the stabilization of the PM phase in $Ni_{50}Mn_{34}In_{15.2}Al_{0.8}$ MSMA. We have also shown that the $T_{PM}$ decreases with increasing magnetic field indicating the role of magnetoelastic coupling. The disappearance of the satellite peaks of the PM phase in the SXRPD pattern in the presence of an external magnetic field provides direct evidence for the crucial role of magnetoelastic coupling in the stabilization of the PM phase in the base and ~5% Al substituted alloy compositions. Our first-principles calculations not only corroborate the experimental findings but also provide a microscopic insight into the role of Al substitution at the In site on the stabilization of PM phase in the ground state of Ni-Mn-In alloys. Our results, thus, put forward Al- substituted Ni-Mn-In MSMA as an ideal system for investigating the physics of precursor phenomena in MSMAs.

## ACKNOWLEDGMENTS

S.S. thanks UGC-DAE CSR, Indore for financial support through "CRS" Scheme. S.S. is thankful to the Science and Engineering Research Board of India for financial support through the award of Ramanujan Fellowship (Grant No: SB/S2/RJN-015/2017) and Early Career Research Award (Grant No: ECR/2017/003186). This work was partially carried out using the facilities of UGC-DAE CSR, Indore. Portions of this research were conducted at the light source PETRA III of DESY, a member of the Helmholtz Association. Financial support from the Department of Science and Technology, Government of India within the framework of the India@DESY is gratefully acknowledged. We would like to thank the beamline scientists Dr. Martin Etter and Dr. Jochi Tseng for their help in setting up the experiments. Authors thank G. Bais and M. Polentarutti of Elettra-Sincrotrone Trieste for their help in the setup for SXRPD measurements under magnetic field. K.K.D thanks DST for providing fellowship through DST- INSPIRE scheme.

*ssingh.mst@iitbhu.ac.in*ssingh.mst@iitbhu.ac.in

# Figures

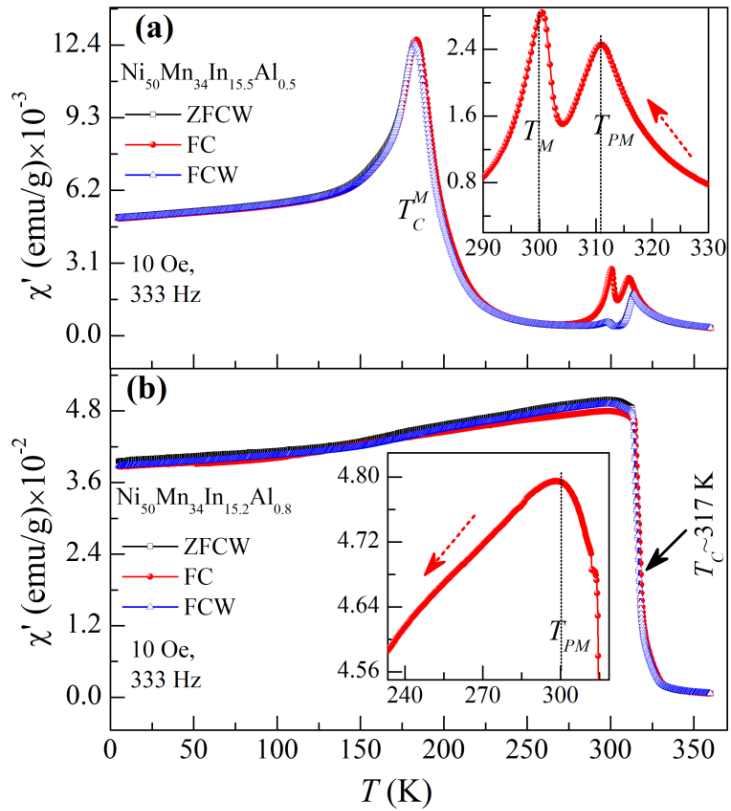

**Fig. 1:** The temperature dependent real part of ac-susceptibility for (a) $Ni_{50}Mn_{34}In_{15.5}Al_{0.5}$ and (b) $Ni_{50}Mn_{34}In_{15.2}Al_{0.8}$ MSMAs. The insets are enlarged view around 300 K for the field cooled protocol. The $T_M$, $T_{PM}$, $T_C^M$ and $T_C$ represent the martensite transition temperature, premartensite

transition temperature, Curie temperature of the martensite phase and Curie temperature of the austenite phase, respectively. The ZFCW, FC and FCW correspond to measurements performed during warming on the zero-field cooled sample, during field cooling and during warming on the field cooled sample, respectively.

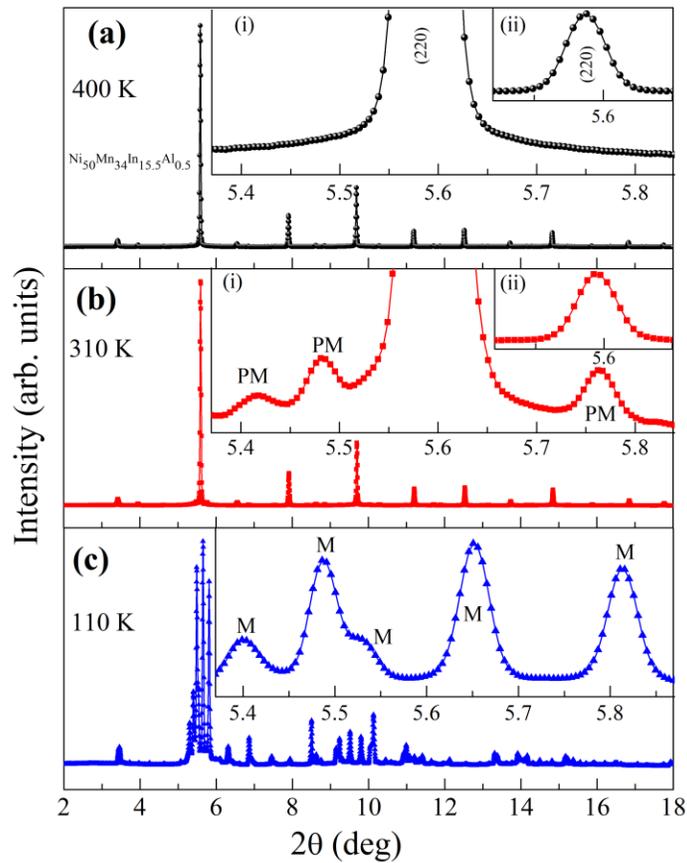

**Fig. 2**: Typical SXRPD patterns of $Ni_{50}Mn_{34}In_{15.5}Al_{0.5}$ MSMA in the **(a)** austenite, **(b)** premartensite and **(c)** martensite phases. An enlarged view around the most intense (220) Bragg peak for the austenite and the premartensite (PM) phases, given in inset (i) of **(a)** and **(b)**, respectively, reveal the appearance of the satellite peaks (indicated by 'PM' in the inset (i) of **(b)**) due to 3*M* like modulation in the PM phase. Untruncated view of the (220) cubic peak for the austenite and PM phases, given in inset (ii) of **(a)** and **(b)**, respectively, reveal the absence of Bain distortion in the PM phase. The inset of **(c)** depicts the splitting of the most intense (220) cubic

peak and appearance of the satellite peaks due to Bain distortion and 3*M* like modulation of the martensite (M) phase. The peaks related to the martensite phase are marked as 'M' in the inset of **(c)**.

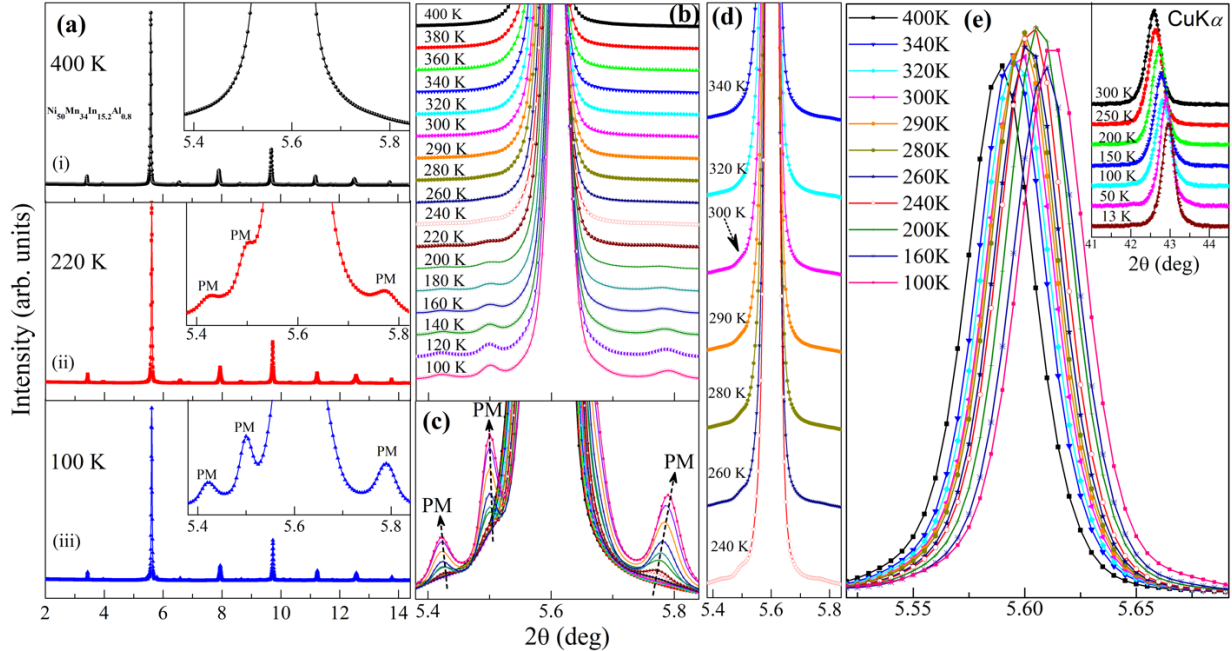

**Fig. 3**: The SXRPD patterns of Ni$_{50}$Mn$_{34}$In$_{15.2}$Al$_{0.8}$ are shown in **(a)** at (i) 400 K, (ii) 220 K and (ii) 100 K. The insets show an enlarged view around the most intense Bragg peak to reveal the satellite peaks of the premartensite (PM) phase. Enlarged view around the most intense cubic peak (220) at various temperature in the range 400-100 K are given in **(b)** and **(c)**. The arrows in **(c)** indicate the temperature dependent shifts of the PM satellite peak positions. Note the gradual sharpening of the satellite peaks in **(c)** on lowering the temperature. **(d)** An enlarged view around the most intense (220) cubic peak at selected temperatures reveal the appearance of the most intense satellite peak of the PM phase at *T* ~ 300 K, indicated by an arrow. **(e)** Untruncated SXRPD profiles of the (220) cubic Bragg peak is depicted in the 400 to100 K range while its inset depicts the XRD profiles in the 300 to 13 K range recorded on a rotating anode-based diffractometer.

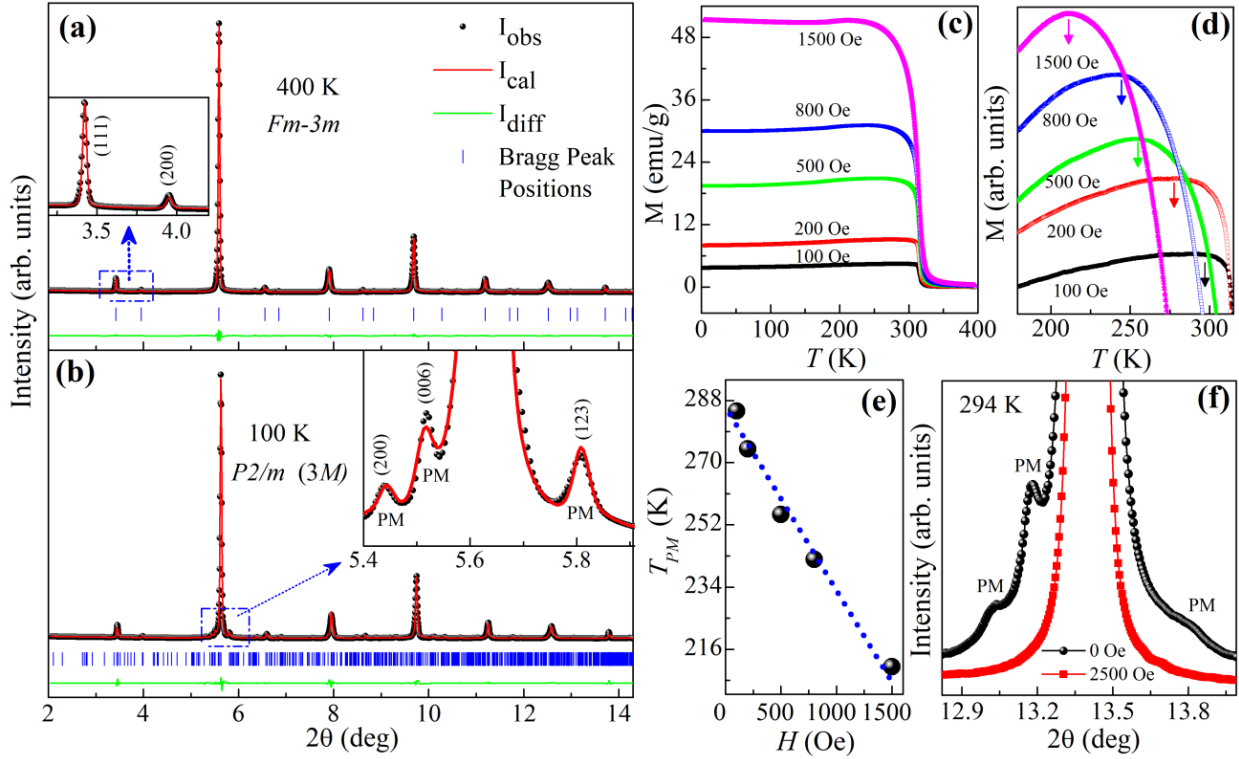

**Fig. 4:** The observed (dark black dots), calculated (continuous red line) and difference patterns (continuous green line) obtained after LeBail refinement using SXRPD pattern of $Ni_{50}Mn_{34}In_{15.2}Al_{0.8}$ MSMA for the **(a)** cubic austenite phase at 400 K and **(b)** 3*M* modulated premartensite (PM) phase at 100 K in the $Fm\bar{3}m$ and *P*2/*m* space groups, respectively. The vertical tick marks above the difference profile represent the Bragg peak positions in **(a)** and **(b)**. The inset of **(a)** shows the presence of (111) and (200) Bragg reflections characteristic of the $L2_1$ ordering in the cubic austenite phase. The inset of **(b)** shows an enlarged view of the fit around the most intense Bragg peak and satellite reflections (marked as 'PM' with their indices) related to the 3*M* modulated PM phase. The temperature dependence of the dc magnetization, measured on zero field cooled sample during warming cycle, is shown in **(c)** for different magnetic fields**.** The enlarged view of **(c)** around the FM $T_C$, shown in **(d),** reveals a skewed diffuse peak due to the PM transition. The variation of the PM transition temperature ($T_{PM}$) with magnetic field is shown in **(e)**. An enlarged view of SXRPD pattern around the most intense Bragg peak collected at 294 K under zero field (black dots connected with continuous line) and an external magnetic field of 2500 Oe (red squares connected with continuous line) is shown in **(f)**. Note the disappearance of the satellite peaks related to the premartensite (PM) phase under the magnetic field.

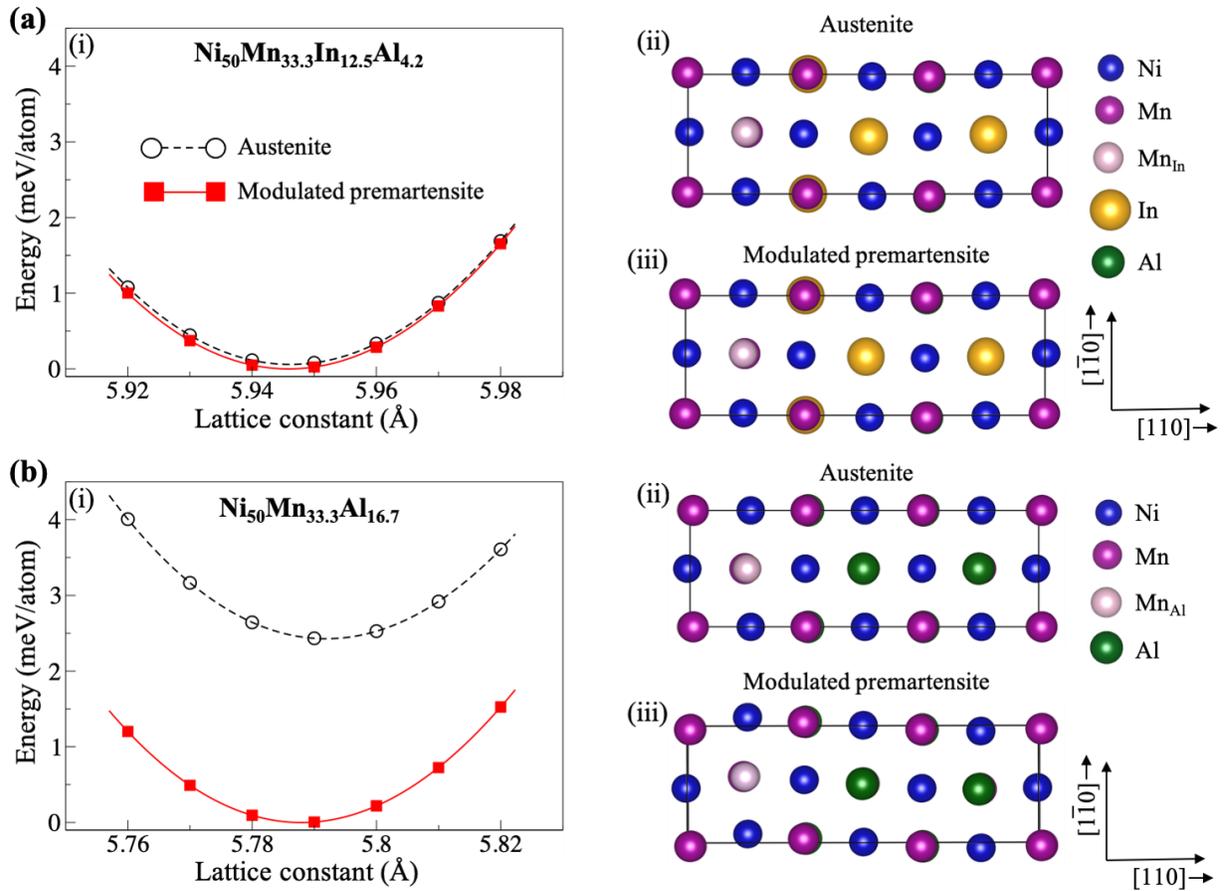

**Fig. 5:** Calculated total energies as a function of lattice constant in the austenite and modulated premartensite phases of (a) $Ni_{50}Mn_{33.3}In_{12.5}Al_{4.2}$ and (b) $Ni_{50}Mn_{33.3}Al_{16.7}$. Supercells with relaxed atomic positions for the austenite and the modulated premartensite phases of $Ni_{50}Mn_{33.3}In_{12.5}Al_{4.2}$ and $Ni_{50}Mn_{33.3}Al_{16.7}$ are shown in insets (ii), (iii) of (a) and (b), respectively. $Mn_{In}$ ($Mn_{Al}$) indicate excess Mn atoms (i.e., above 25% Mn content) in our simulation cells occupying In (Al) sublattice. The Al atoms are behind the In atoms in the insets (ii) and (iii) of (a).

# Supplemental Material

# Robust evidence for the stabilization of the premartensite phase in Ni-Mn-In magnetic shape memory alloys by chemical pressure

Anupam K. Singh,[1] Sanjay Singh,[1]* B. Dutta,[2] K. K. Dubey,[1] Boby Joseph,[3] R. Rawat,[4] and Dhananjai Pandey[1]

[1]*School of Materials Science and Technology, Indian Institute of Technology (Banaras Hindu University) Varanasi-221005, India.*
[2]*Department of Materials Science and Engineering, Faculty of Mechanical, Maritime and Materials Engineering, Delft University of Technology, Mekelweg 2, 2628 CD Delft, The Netherlands*
[3]*Elettra-Sincrotrone Trieste, S. C. p. A. S.S.14, Km 163.5 in Area Science Park, Basovizza 34149, Italy*
[4]*UGC-DAE Consortium for Scientific Research, University Campus, Khandwa Road, Indore 452001, India*

**Experimental and Computational Methods:**

The polycrystalline alloys with ~3% and 5% Al substitution in place of In site in $Ni_{50}Mn_{34}In_{16}$ leading to nominal compositions $Ni_{50}Mn_{34}In_{15.5}Al_{0.5}$ and $Ni_{50}Mn_{34}In_{15.2}Al_{0.8}$, respectively, were prepared by the conventional arc-melting technique[1] under argon atmosphere using the appropriate quantity of each constituent element (99.99% purity). The samples were melted several times to get good homogeneity. An extra 2% manganese (Mn) was added before melting to compensate for the Mn loss due to its evaporation during melting. The melt-cast ingot was annealed in vacuum-sealed quartz ampoule at 800ºC for 24 h to achieve further homogeneity and then quenched in ice water. A part of the homogenized bulk sample was crushed into powder using a mortar pestle and sealed in quartz ampoule under argon atmosphere followed by annealing at 500ºC for 12 h and finally furnace cooled to remove residual stresses[2, 3], if any, introduced during the grinding. We used this stress free annealed powder sample for all the characterizations. The chemical composition was checked using the energy dispersive analysis of x-ray (EDAX) technique. The compositions so obtained are $Ni_{50.05}Mn_{33.93}In_{15.57}Al_{0.53}$ and $Ni_{51.9}Mn_{34.42}In_{12.76}Al_{0.81}$ which correspond to $Ni_{50}Mn_{34}In_{15.5}Al_{0.5}$ (or $Ni_2Mn_{1.36}In_{0.62}Al_{0.02}$) and $Ni_{50}Mn_{34}In_{15.2}Al_{0.8}$ (or $Ni_2Mn_{1.36}In_{0.61}Al_{0.03}$), respectively.

X-ray powder diffraction (XRD) data were collected using an 18-kW rotating Cu anode-based x-ray diffractometer (Rigaku) fitted with a curved graphite crystal monochromator in the diffraction beam. A closed-cycle He refrigerator-based low temperature attachment was used for XRD measurements in the 300 to 13 K range. In addition, high-resolution synchrotron x-ray powder diffraction (SXRPD) data were collected in the 400-100 K range at temperature interval of 10 K in the cooling cycle at a wavelength of $\lambda \sim 0.207$ Å at P02.1 beamline of Petra-III DESY, Germany. The sample containing capillary was spinning continuously to minimize the texturing effect in the SXRPD data. Further, high-resolution SXRPD data were collected without and with magnetic field bias (0 Oe and 2500 Oe) at 294 K at a wavelength of $\lambda \sim 0.495$ Å on the Xpress beamline of Elettra, Italy[4]. The sample containing capillary was oscillated continuously to minimize the texturing effect in the SXRPD data collected at Elettra.

The temperature dependent ac-susceptibility ($\chi'(T)$) at a drive field 10 Oe for 333.33 Hz frequency and dc magnetisation ($M(T)$) data at 100 Oe were collected under zero field cooled warming (ZFCW), field cooled (FC), and field cooled warming (FCW) protocols using a superconducting quantum interference device (SQUID) based magnetometer (Quantum Design, MPMPS-3). For the ZFCW protocol, the sample was cooled from 380 K (well above its ferromagnetic $T_C$) down to 5 K in the absence of magnetic field and then the $\chi'(T)$ and $M(T)$ data were collected up to 360 K during warming. Further, the data were also collected while cooling the sample under field (FC) and during the warming cycle on the field cooled sample (FCW). The magnetic field dependent isothermal magnetisation ($M(H)$) and $M(T)$ at several fields under ZFCW protocol were collected using a vibrating sample magnetometer (VSM) module equipped in a physical properties measurement system (PPMS) (Dynacool, Quantum Design). The resistivity data was collected in the temperature range of 8 to 300 K in the warming cycle using Cryogen Free Measurement System (Cryogenic Limited, UK).

First-principles total energy calculations were performed based on spin-polarized Density Functional Theory (DFT) as implemented in the Vienna *ab initio* simulation package (VASP)[5-7]. In VASP calculations, 24-atom supercells were used to model the austenite and the modulated premartensite (PM) phases for three different atomic compositions: $Ni_{50}Mn_{33.3}In_{16.7}$, $Ni_{50}Mn_{33.3}In_{12.5}Al_{4.2}$ and $Ni_{50}Mn_{33.3}Al_{16.7}$. The plane-wave basis projector augmented wave (PAW) method[8] together with the generalized gradient approximation (GGA) for the exchange-correlation potential as parameterized by Perdew, Burke and Ernzerhof (PBE)[9] were used for obtaining relaxed geometries and atomic positions. An energy cut-off of 450 eV was chosen for the plane-wave basis and a Monkhorst-Pack scheme[10] with a k-point grid of 6×18×16 was used for the Brillouin zone sampling. Convergence criterion of $10^{-7}$ eV was used for the self-consistent electronic loop. All atomic positions were relaxed until the residual forces acting on the atoms were less than $10^{-3}$ ev/Å.

**Crystal structure of Ni-Mn-In-Al magnetic shape memory alloys (MSMAs):**

Phase purity and crystal structures were confirmed by x-ray powder diffraction measurements. The average long-range ordered structure was confirmed by LeBail technique using the high-resolution synchrotron as well as laboratory source x-ray powder diffraction patterns. All the refinements were carried out using FULLPROF package[11]. For the Al ~ 3% composition with chemical formula $Ni_{50}Mn_{34}In_{15.5}Al_{0.5}$, the LeBail refinement using the SXRPD data confirms the cubic austenite phase structure in the $Fm\bar{3}m$ space group at 400 K. The observed, calculated and difference profiles obtained after the refinement are shown in Fig. S1(a). The refined lattice parameter is found to be 6.01009(6) Å. The presence of (111) and (200) Bragg reflections (the inset (i) of Fig. S1(a)) confirms the L2$_1$ ordering in the cubic austenite phase[12]. The LeBail refinement using SXRPD data at 310 K reveals that all the peaks can be indexed using the 3$M$ modulated monoclinic structure in the $P2/m$ space group with $a$ = 4.3869(7) Å, $b$ = 5.6866(1) Å, $c$ = 13.0028(2) Å and $\beta$ = 93.695(3)° as can be seen from the excellent fit between the observed and calculated profiles shown in Fig. S1(b). The LeBail refinement using the SXRPD data at 110 K reveals that all the peaks can be indexed using 3$M$ modulated monoclinic structure in the $P2/m$ space group as can be seen from Fig. S1(c), which depicts the fit between the observed and calculated profiles. The LeBail refinement using the laboratory source XRD data also shows the 3$M$ modulated monoclinic structure in the $P2/m$ space group at 300 K (see Fig. S2(a)). The laboratory source XRD data at different temperatures (300 to 13 K) are shown in Fig. S2(b). The enlarged view around most intense Bragg peak, shown in the inset of Fig. S2(b), confirms the 3M modulated monoclinic martensite phase ($P2/m$ space group) upto the lowest temperature (13 K) up to data was collected for Ni$_{50}$Mn$_{34}$In$_{15.5}$Al$_{0.5}$ MSMA.

For the second composition (Al ~ 5%) with the chemical formula Ni$_{50}$Mn$_{34}$In$_{15.2}$Al$_{0.8}$, the LeBail refinement using of laboratory source XRD data at 300 K confirms the cubic in $Fm\bar{3}m$ space group symmetry of the austenite phase. The results of the refinements are shown in Fig. S3(a). The refined lattice parameter is found to be 6.0135(2) Å. The inset of Fig. S3(a) shows the presence of (111) and (200) Bragg reflections which confirms the L2$_1$ ordering in the austenite phase. The laboratory source XRD data at different temperatures from 300 K to 13 K are shown in Fig. S3(b). The enlarged view around the most intense Bragg peak is shown in the inset of Fig. S3(b), which reveals the absence of any Bain distortion as the cubic peaks do not exhibit any splitting down to 13 K up to which the data was collected for $Ni_{50}Mn_{34}In_{15.2}Al_{0.8}$ MSMA.

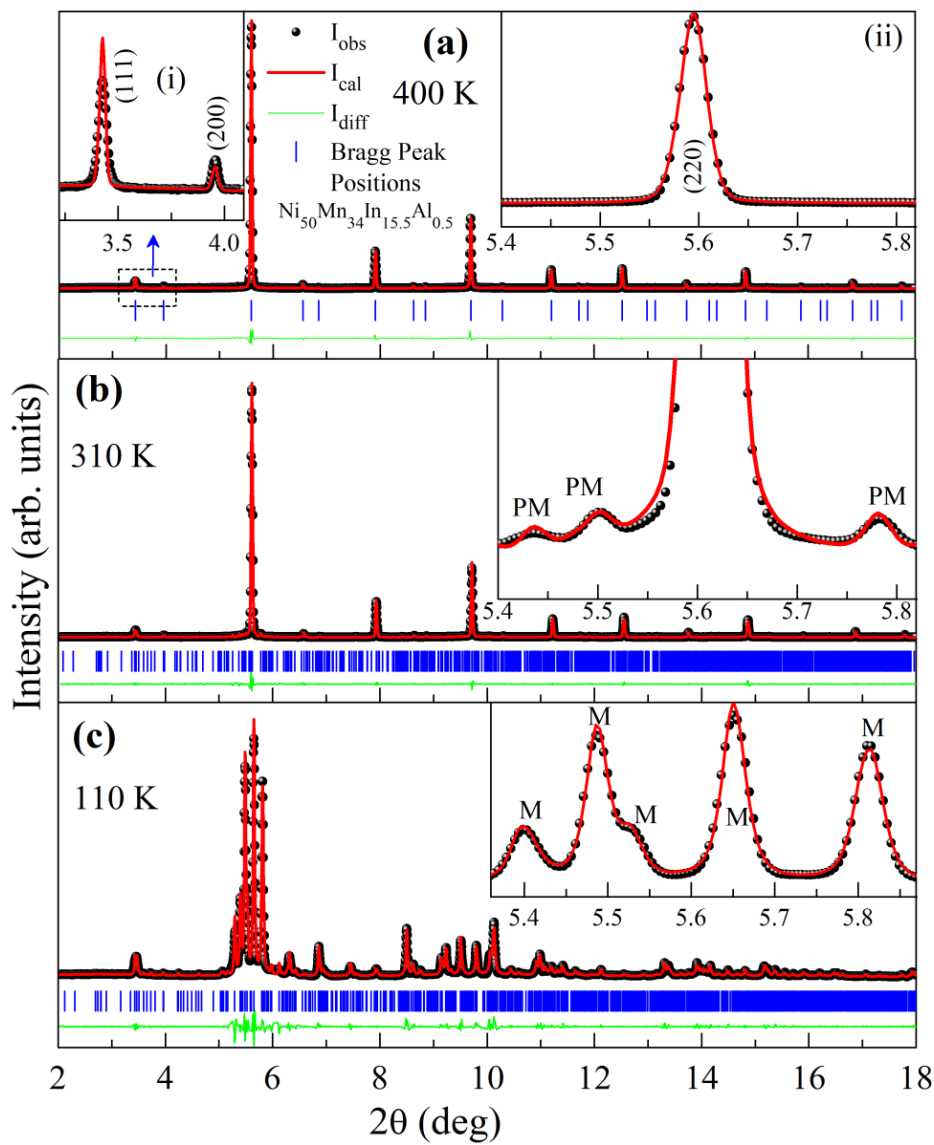

**Fig. S1**: The observed (dark black dots), calculated (continuous red line) and difference patterns (continuous green line) obtained after LeBail refinement using the SXRPD data for the **(a)** cubic austenite, **(b)** 3*M* modulated premartensite (PM) and **(c)** 3*M* modulated martensite phases in the $Fm\bar{3}m$, *P2/m* and *P2/m* space groups, respectively, for $Ni_{50}Mn_{34}In_{15.5}Al_{0.5}$. The vertical tick marks above the difference profile represent the Bragg peak positions. The insets (i) and (ii) of **(a)** show an enlarged view of fit around the (111) and (200) Bragg reflections and around the most intense

Bragg peak, respectively. The inset of **(b)** and **(c)** shows fits around the most intense Bragg peak region in a magnified scale. The satellite peaks of the PM phase are marked as 'PM' at 310 K in the inset of **(b)**. The peaks related to the martensite phase are marked as 'M' at 110 K in the inset of **(c)**.

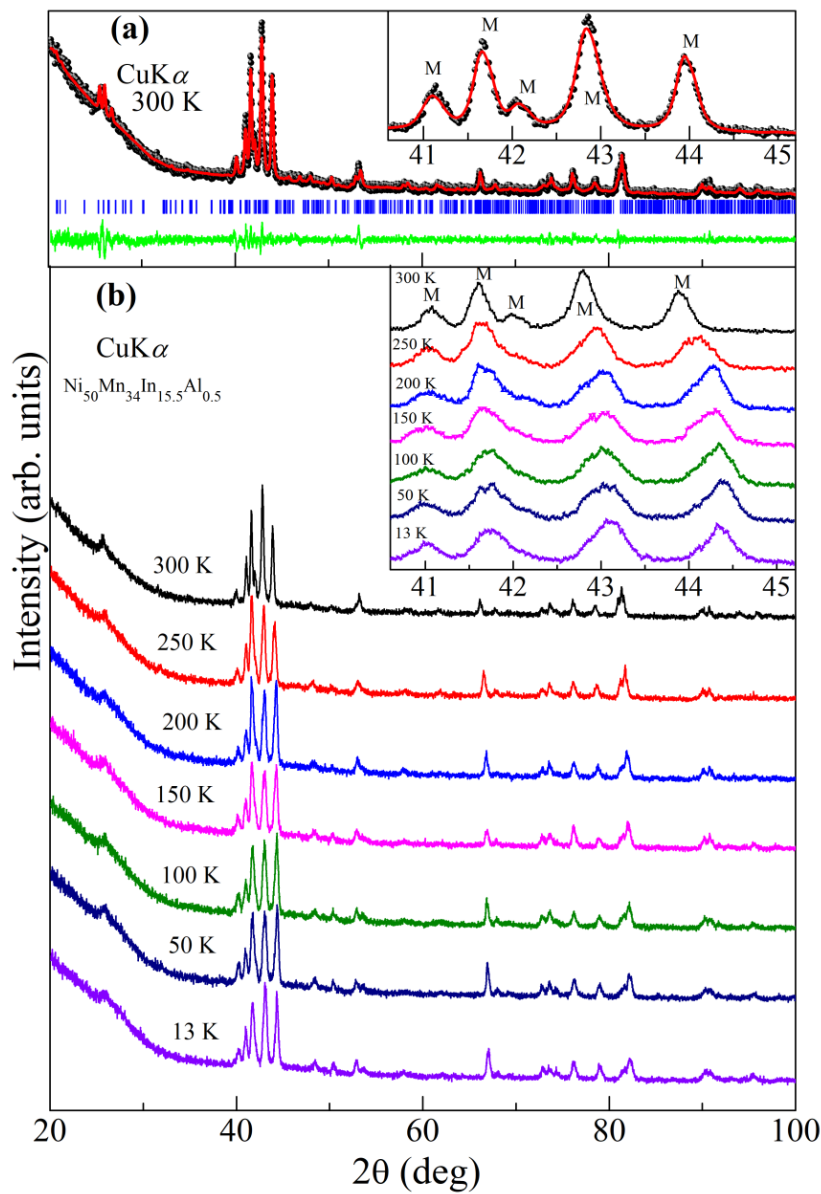

**Fig. S2**: **(a)** The observed (dark black dots), calculated (continuous red line) and difference profiles (continuous green line) obtained after LeBail refinement using laboratory source XRD data at 300 K for the martensite phase in the *P2/m* space group for $Ni_{50}Mn_{34}In_{15.5}Al_{0.5}$. The vertical tick marks above the difference profile represent the Bragg peak positions. The inset of **(a)** shows enlarged

view of fit around most intense Bragg peak. The laboratory source XRD data at indicated temperature (300 to 13 K) are shown in **(b)** where the inset shows enlarged view around most intense Bragg peak region. The peaks related to the martensite phase are marked as 'M' in the inset of (a) and (b).

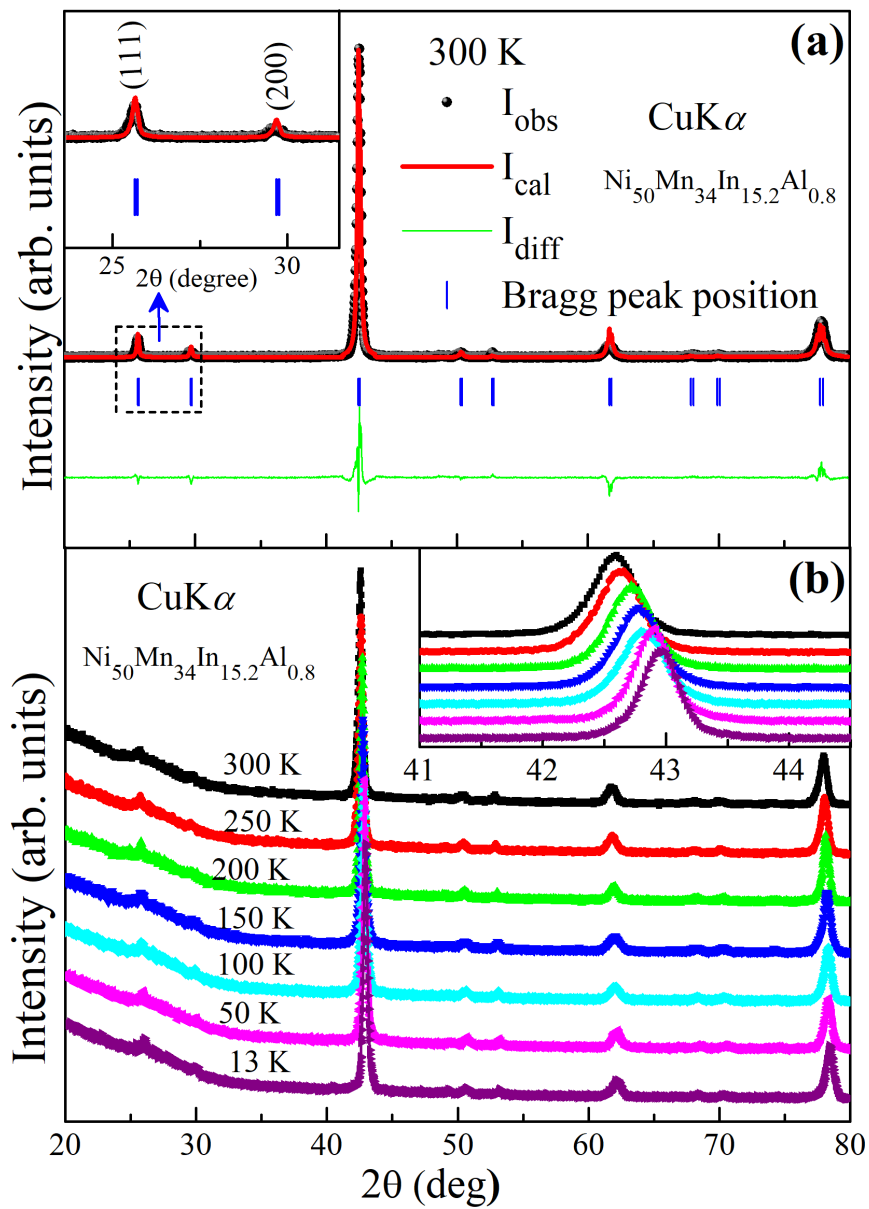

**Fig. S3**: **(a)** The observed (dark black dots), calculated (continuous red line) and difference profiles (continuous green line) obtained after LeBail refinement using laboratory source XRD data at 300 K for the cubic austenite phase in the $Fm\bar{3}m$ space group for $Ni_{50}Mn_{34}In_{15.2}Al_{0.8}$. The vertical tick

marks above the difference profile represent the Bragg peak positions. The inset of **(a)** shows enlarged view of fit around the (111) and (200) Bragg reflections. The laboratory source XRD data are shown in **(b)** at indicated temperatures (300 to 13 K) where inset shows enlarged view around the most intense Bragg peak.

**High-resolution SXRPD data for $Ni_{50}Mn_{34}In_{15.2}Al_{0.8}$ MSMA under magnetic field:**

High-resolution SXRPD patterns was recorded at $\lambda \sim 0.495$ Å without and in the presence of magnetic field (0 Oe and 2500 Oe) at 294 K for $Ni_{50}Mn_{34}In_{15.2}Al_{0.8}$ MSMA. The results are shown in Fig. S4(a) and (b). A simple custom-made setup, based on a permanent magnet giving a field of ~2500 Oe at the sample, fitted to the beamline sample stage with the sample containing capillary in the center, was used for applying the magnetic field. A photograph depicting the setup is shown in Fig. S4(c). The inset of Fig. 4(a) depicts an enlarged view of lower $2\theta$-range. The '*' marked peaks in inset of Fig. S4(a) are due to the setup used for measurement under magnetic field (shown in Fig. S4(c)). An enlarged view of encircled region of Fig. S4(a) around the most intense Bragg peak is shown in Fig. S4(b). It reveals the satellite peaks related to the PM phase, labeled as 'PM', for zero field (H = 0 Oe). It is evident from the figure that the satellite peaks related to the PM phase disappear completely in the presence of 2500 Oe magnetic field. This provides direct evidence for the role of the magnetoelastic coupling on the stability of the PM phase in $Ni_{50}Mn_{34}In_{15.2}Al_{0.8}$. Our initial results with and without magnetic field calls for further such systematic measurements as a function of temperature to obtain further inputs on magnetoelastic coupling in this alloy system.

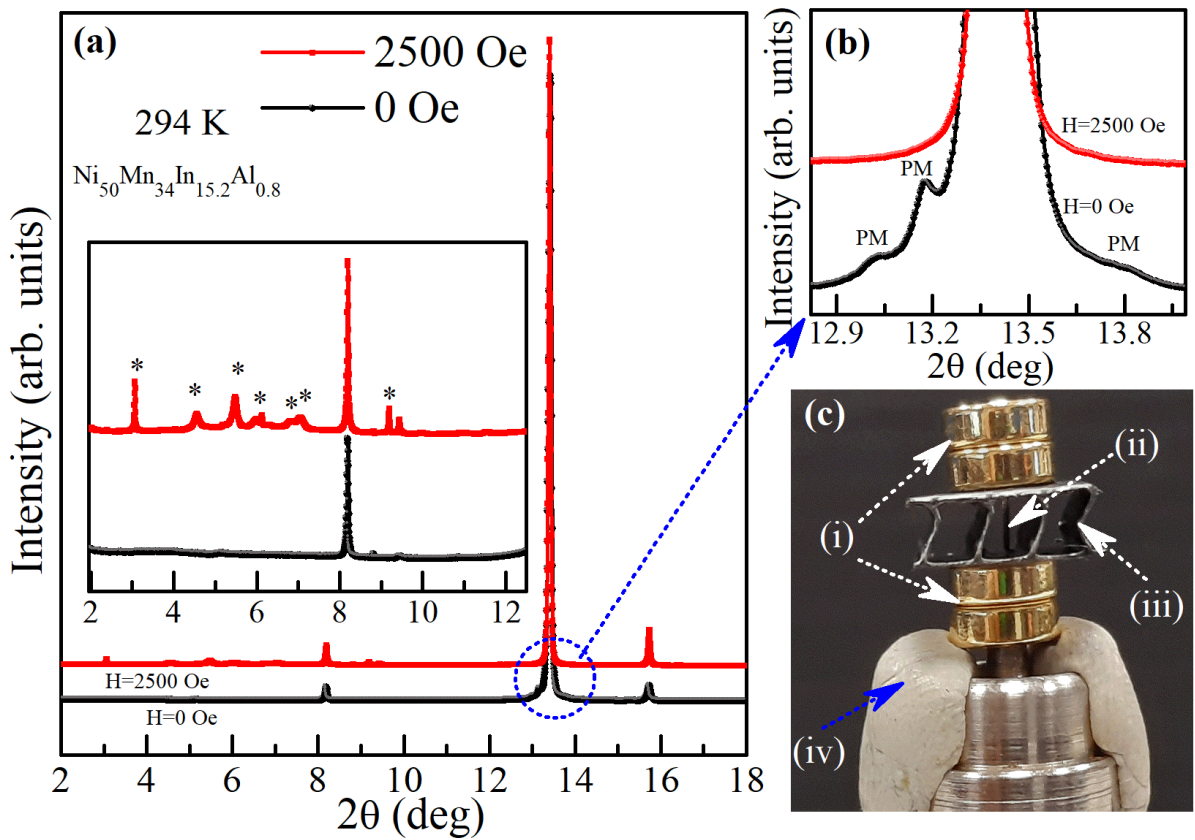

**Fig. S4: (a)** High-resolution SXRPD pattern ($\lambda \sim 0.495$ Å) without and with the presence of magnetic field (0 Oe and 2500 Oe) at 294 K for $Ni_{50}Mn_{34}In_{15.2}Al_{0.8}$. The inset of **(a)** shows an enlarged view of the lower 2θ-range (2-12.5°). **(b)** An enlarged view of encircled region of **(a)** (guided by arrow). **(c)** Image of magnets setup used in measurement, where (i), (ii), (iii) and (iv) indicate the magnets, capillary position, plastic for support, and clay for positioning the magnets centered with respect to the capillary, respectively (shown by arrow). The '*' in the inset of **(a)** and 'PM' in **(b)** indicate the extra peaks related to the setup of magnets and satellite peaks related to the PM phase, respectively.

**Magnetisation:**

The temperature dependent dc magnetisation plots measured at 100 Oe under ZFCW, FC and FCW protocols are shown in Fig. S5(a) for $Ni_{50}Mn_{34}In_{15.2}Al_{0.8}$ MSMA. The field dependent isothermal magnetisation at 2 K is shown in Fig. S5(b), which reveals a high value of saturation magnetisation ($M_S \sim 5.82\ \mu_B/f.u.$) for $Ni_{50}Mn_{34}In_{15.2}Al_{0.8}$ MSMA.

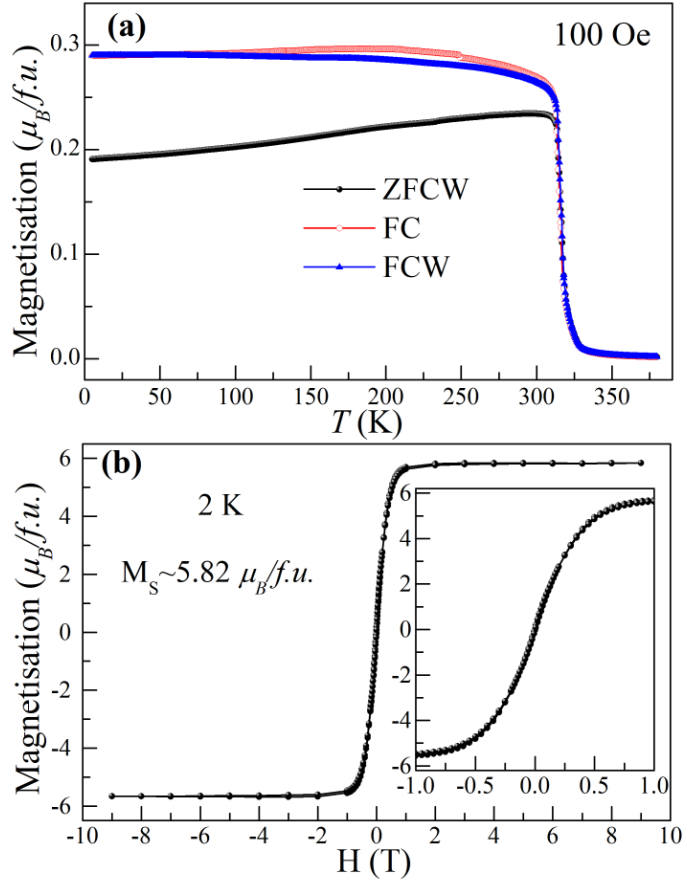

**Fig. S5**: **(a)** The temperature dependency of dc magnetisation behavior measured at 100 Oe under ZFCW, FC and FCW protocols for $Ni_{50}Mn_{34}In_{15.2}Al_{0.8}$. **(b)** The magnetic field dependent magnetisation (M-H loop) at 2 K for $Ni_{50}Mn_{34}In_{15.2}Al_{0.8}$. The inset of **(b)** shows enlarged view around ± 1 Tesla.

**Resistivity:**

The resistivity of $Ni_{50}Mn_{34}In_{15.2}Al_{0.8}$ as a function of temperature measured during the warming cycle is shown in Fig. S6. It shows smooth monotonic behavior without any step or peak expected for a martensite transition. This feature confirms the absence of martensite transition in $Ni_{50}Mn_{34}In_{15.2}Al_{0.8}$.

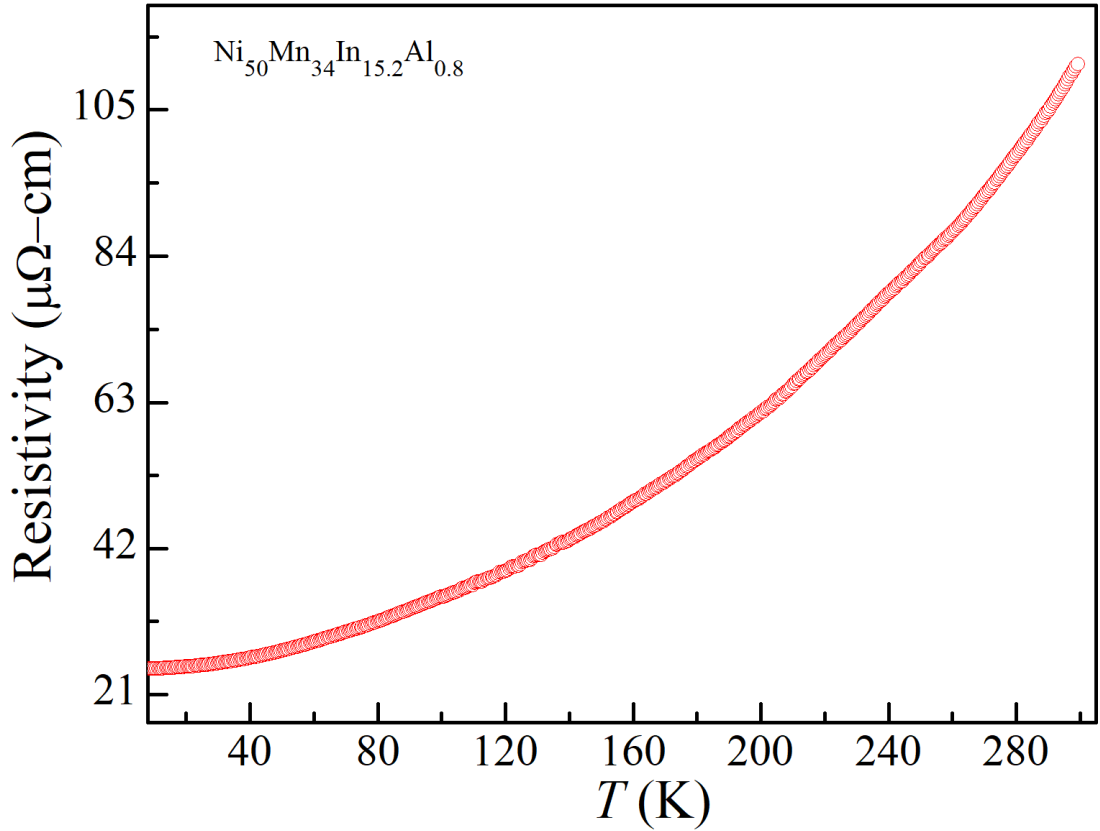

**Fig. S6:** Temperature dependency of resistivity measured during warming cycle for $Ni_{50}Mn_{34}In_{15.2}Al_{0.8}$.

**First- principles total energy calculations in the ferromagnetic state of $Ni_{50}Mn_{33.3}Al_{16.7}$:**

The origin of the PM phase in Al substituted Ni-Mn-In alloy, investigated in the present manuscript, lies in the coupling between magnetism and crystal structure, i.e., magnetoelastic coupling. The effect of magnetism on the stability of the PM phase can be qualitatively understood from the total energies calculated at two different magnetic states, i.e., antiferromagnetic and ferromagnetic states. In order to highlight the effect of magnetism on the stability of the PM phase concretely, we have, in addition to the antiferromagnetic state discussed in the manuscript (inset (i) of Fig. 5(b) in the manuscript), chosen an artificial ferromagnetic state in $Ni_{50}Mn_{33.3}Al_{16.7}$ with all Mn spins oriented in the same direction and calculated total energies of both austenite and PM phase. Our calculations in the ferromagnetic state with significantly higher magnetic moment per atom than the antiferromagnetic state reveal almost negligible atomic modulation amplitude for the PM phase resulting in tiny energy difference (approximately 0.1 meV/atom) between the austenite and the PM phase (see Fig. S7). This indicates that a high magnetic moment resists the formation of the modulated PM phase and this phase may not be formed in the ferromagnetic state of $Ni_{50}Mn_{33.3}Al_{16.7}$.

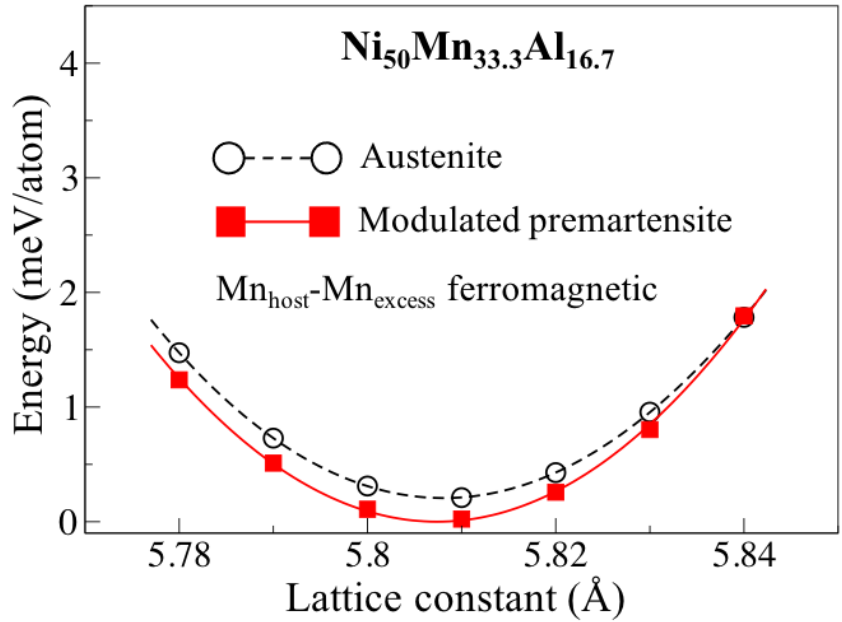

**Fig. S7:** Calculated total energies as a function of lattice constant for the austenite and the premartensite phase of $Ni_{50}Mn_{33.3}Al_{16.7}$ in the ferromagnetic state.